\documentclass[a4paper,10pt,twoside]{cpc-hepnp}

\usepackage{multicol}
\usepackage{graphicx}
\usepackage{booktabs}
\usepackage{amssymb,bm,mathrsfs,bbm,amscd}
\usepackage[tbtags]{amsmath}
\usepackage{lastpage}

\begin{document}

\fancyhead[co]{\footnotesize YAN Xin-Hu~ et al:NaI (Tl) Calorimeter Calibration and Simulation for Coulomb Sum Rule Experiment in Hall-A at Jefferson Lab }


\title{NaI (Tl) Calorimeter Calibration and Simulation  for Coulomb Sum Rule Experiment in Hall-A at Jefferson Lab\thanks{Supported
    by the National Natural Science Foundation of China(10605022,10875053) and US department of Energy contract DE-AC05-84ER-40150 under which Jefferson Science Associates operates the Thomas Jefferson National Accelerator Facility
}}

\author{%
       YAN Xin-Hu$^{1;1)}$\email{yanxinhu@ mail.ustc.edu.cn}%
\quad C. Alexandre$^{2}$ 
\quad CHEN Jian-Ping$^{2}$
\quad C. Seonho$^{4}$ 
\quad LV Hai-Jiang$^{5}$\\
\quad M. Zein-Eddine$^{3}$ 
\quad Oh Yoomin$^{4}$  
\quad S. Vincent$^{2}$ 
\quad YE Yun-Xiu$^{1}$ 
\quad YAO Huan$^{3}$  
}
\maketitle

\address{%
$^1$Department of Modern Physics, University of Science and Technology of China, Hefei, 230026, China\\
$^2$Thomas Jefferson National Accelerator Facility,Newport News,VA 23606, USA)\\
$^3$Temple University, Philadelphia, PA 19122, USA\\
$^4$Seoul National University, Seoul 151-747 Korea\\
$^5$Huangshan University, Huangshan, Anhui, 245021, China\\

}

\begin{abstract}
 A precision measurment of inclusive electron scattering cross sections was carried out at Jefferson Lab in the quasi-elastic region for ${^4}$He, ${^{12}}$C, ${^{56}}$Fe and ${^{208}}$Pb targets. Longitudinal ($R_{L}$) and transverse ($R_{T}$) response functions of nucleon were extracted in the momentum transfer range 0.55 GeV/c$\le$$|q|$$\le$1.0 GeV/c. To achieve the above goal, a NaI (Tl) calorimeter was used to distinguish good electrons from background including pions and low energy electrons rescattered from walls of the spectrometer magnets. Due to a large set of  kinematics and changes in HV settings, a number of calibrations were performed for the NaI (Tl) detector. Corrections for a few blocks of NaI (Tl) with bad or no signal were applied. The resolution of NaI (Tl) detector after calibration reached $\frac{\delta E}{\sqrt{E}} \approx 3\%$ at E=1 GeV. The performance of NaI (Tl) detector was compared with a simulation.

\end{abstract}



\begin{multicols}{2}

\section{Introduction}
 The study of nucleon properties in a nuclear medium is an essential objective in nuclear physics. The Coulomb Sum Rule(CSR) provides one of the cleanest means to study nuclear medium effects on the charge response of the nucleons\cite{morgenstern}. The Coulomb sum $S_{L}(q)$ is given by 
\begin{equation}
S_L(q) =\frac{1}{Z}\int_{\omega^{+}}^\infty d\omega
\frac{R_L(q,\omega)}{\tilde{G}_{E}(Q^{2})^{2}} \,,
\end{equation}
with $Z$ the atomic number of the nucleus, $Q^2$ the four momentum transfer squared, $q$ the three momentum transfer and $\omega$ the energy loss. After factoring out an effective nucleon charge form factor $\tilde{G}_{E}(Q^{2})^{2}$ which is an appropriate sum of neutron and proton charge form factors and the longitudinal response $R_{L}(q,\omega)$ is integrated from $\omega^{+}$ to infinity where $\omega^{+}$ is selected to exclude the elastic peak, the $S_{L}(q)$ should approach 1 as $q\rightarrow \infty$ for a system of non-relativistic nucleon. In this limit $S_{L}(q)$ simply measures the total charge divided by Z. In the Fermi gas model the asymptotic limit of $S_{L}(q)$ is reached for $q\ge 2k_{F}\sim $ 500 MeV/c where correlations due to the Pauli Blocking effect vanish. Since the ratio of $R_{L}$ to $R_{T}$ is small at large $q$, $R_{L}$ has a large sensitivity to the uncertainties of the cross sections. To make a precision measurement of $R_{L}$, the uncertainties of cross sections are needed to be at 1\% level. Therefore, $R_{L}$ was much harder to determine with good precision than $R_{T}$. 
\vspace{2mm}

   Jefferson Lab CSR experiment (E05-110)\cite{pro} measured the cross sections of quasi-elastic electron scattering on four different targets ( ${^4}$He, ${^{12}}$C, ${^{56}}$Fe and ${^{208}}$Pb) at four different scattering angles ($15^{\circ}$, $60^{\circ}$, $90^{\circ}$, $120^{\circ}$) with beam energies from 0.4 GeV to 4.0 GeV. The standard Hall-A detector configuration includes two high resolution spectrometers (HRS)\cite{Al}. Each HRS has a $Q_{1}Q_{2}DQ_{3}$ magnet configuration where Q1, Q2 and Q3 are quadrupole magnets and D is a dipole magnet. For the CSR experiment, both HRSs were configured for electron detection. The NaI (Tl) detector was installed in the left HRS. We will focus on the left HRS. The left HRS detector package consisted of two Vertical Drift Chambers (VDCs), a pair of plastic scintillator planes, a gas Cerenkov counter and a NaI (Tl) calorimeter. The VDCs were used to determine particle trajectory. The scintillators made the trigger. The gas Cerenkov counter and the NaI (Tl) calorimetor formed the particle identification (PID) system.
    
\section{NaI (Tl) Detector}
 This NaI (Tl) calorimeter was first used at Los Alamos National Laboratory\cite{NaINIM} and Brookhaven National Laboratory. The detector was transferred to Jefferson Lab for this experiment. The NaI (Tl) calorimeter was refurbished and reconfigured into three boxes with each box consisting of 90 (10$\times$9 array) blocks. The length, width and height of each individual block are 30.5cm, 6.35cm and 6.35cm, respectively. Because the total length of 30.5 cm was 11.5 radiation length, an electron with less than 0.55 GeV could deposit most of its energy in the calorimeter. An electron with energy higher than 0.55 GeV would have some energy leaked. Since a few blocks have bad or no signal during the experiment, the missing energy corrections for the bad blocks were studied and corrected in the calibration. The following section will focus on the middle box of the NaI (Tl) detector.

    \section{NaI (Tl) Calibration}
     An electromagnetic cascade generates a shower of low energy photons and electron-positron pairs when a high energy electron hits the NaI (Tl) calorimeter. As the cascade propagates, a large part of the original particle's energy is converted to lights, which usuall covers several blocks. The light in each block is collected in a photomultiplier tube (PMT). The output signal from the PMT is then digitized with an analog-to-digital-converter (ADC). There is a conversion between raw ADC values and the total energy deposition in the NaI (Tl) calorimeter. To accomplish this conversion, one needs to determine calibration coefficients for each of the calorimeter blocks. In general the electromagnetic cascade is spread over several adjacent blocks, the output signal must be integrated over the entire calorimeter volume to obtain the total detectable energy. If the calorimeter has been calibrated properly, the total deposited energy E should be proportional to the incident particle's energy (or momentum $p$). 
     
     \subsection{Calibration event selection }
     To obtain good calibration coefficients, an electron sample needed to be selected. This was accomplished by selecting a run with less background from pions. Runs in the quasi-elastic electron scattering settings were used to do the calibration. The following tight cuts were applied to select electron sample:\\   
     (1) An event reconstruction in the spectrometer detector package was successful;
     (2) Only one track had been reconstructed by the VDC system;
     (3) The event was identified as an electron with a tight Gas Cerenkov cut;
     (4) The event was in a good acceptance region of spectrometer;
     (5) The event was in the central region of the block being calibrated.

    \subsection{The method for determination of NaI (Tl) calibration coefficients}
   The calibration coefficients are defined to transform the ADC amplitude of each block into the energy deposition of the electron in this block. Since the Moliere radius of NaI (Tl) is 4.8 cm\cite{zhu}, the incident particle's energy is deposited in the 9 blocks when it hits the central block (i.e. blk5 in Fig. 1). The basic calibration cell is set to be 9 blocks. A linear minimization method is used to determine the calibration coefficients. The Chi-square minimization function is defined as follows:   
    \begin{equation}
      \chi^2= \sum_{j=0}^{N}\left(E_{kin}^{j}-\sum_{k=0}^{9}C_{k}A_{k}^{j}\right)^2 \,,
    \end{equation}
    where $j$ is the index of the selected calibration events and $k$ is the index of the NaI (Tl) blocks. $A_{k}^{j}$ is the amplitude in the $k$-th NaI (Tl) block for the $j$-th event. $E_{kin}$ is the scattering electron energy; $C_{k}$ is the calibration constant for the $k$-th block.
    \begin{equation}
      \frac{\partial}{\partial C_{i}}\chi^2=\frac{\partial}{\partial C_{i}}\sum_{j=0}^{N}\left(E_{kin}^{j}-\sum_{k=0}^{9}C_{k}A_{k}^{j}\right)^2\,,
    \end{equation}
   where $i$ varies between 0 and 90. $\chi^2$ is minimized when the above quantity is set to zero. It leads to: 
    \begin{equation}
      \sum_{k=0}^{9}\left(C_{k}\left(\sum_{j=0}^{N}A_{k}^{j}A_{i}^{j}\right)\right)=\sum_{j=0}^{N}E_{kin}^{j}A_{i}^{j}\,,
    \end{equation}
    The linear equation can be summarized in matrix form ...

    \begin{center}
      \begin{equation}
      MC=E \,,
      \end{equation}
    \end{center}
   where C and E are defined as vectors\\

    \begin{center}
      \begin{equation}
        C=\left(\begin{array}{c} C_{0}\\.\\.\\. \\C_{n}\end{array}\right)\,,
      \end{equation}
    \end{center}

    \begin{center}
      \begin{equation}
        E=\left(\begin{array}{c} \sum_{j=0}^{N}E_{kin}^{j}A_{0}^{j}\\.\\.\\. \\ \sum_{j=0}^{N}E_{kin}^{j}A_{n}^{j}\end{array}\right)\,,
      \end{equation}
    \end{center}

    and the matrix elements are given by:
    \begin{center}
      \begin{equation}
        M_{ij}=\sum_{k=0}^{N}A_{i}^{k}A_{j}^{k}\,,
      \end{equation}
    \end{center}

   At the end, the calibration coefficients are obtained by inverting eq. (5):
    \begin{center}
      \begin{equation}
        C=M^{-1}E \,,
      \end{equation}
    \end{center}
    \subsection{Missing energy correction for bad blocks}
    To find an average value to correct for the missed energy in the neighboring block which was bad, 9 good blocks were selected for studying the amplitude ratio of adjacent block to the central one. After obtaining the relationship, the missing energy of bad blocks could be corrected back for calibration. A few small circle cuts in the region of central block were set to obtain the amplitude ratio for scattering electron with p= 539 MeV/c at $60^{\circ}$.

    \begin{center}
      \vspace{1mm}
      \includegraphics[width=6cm,height=6cm]{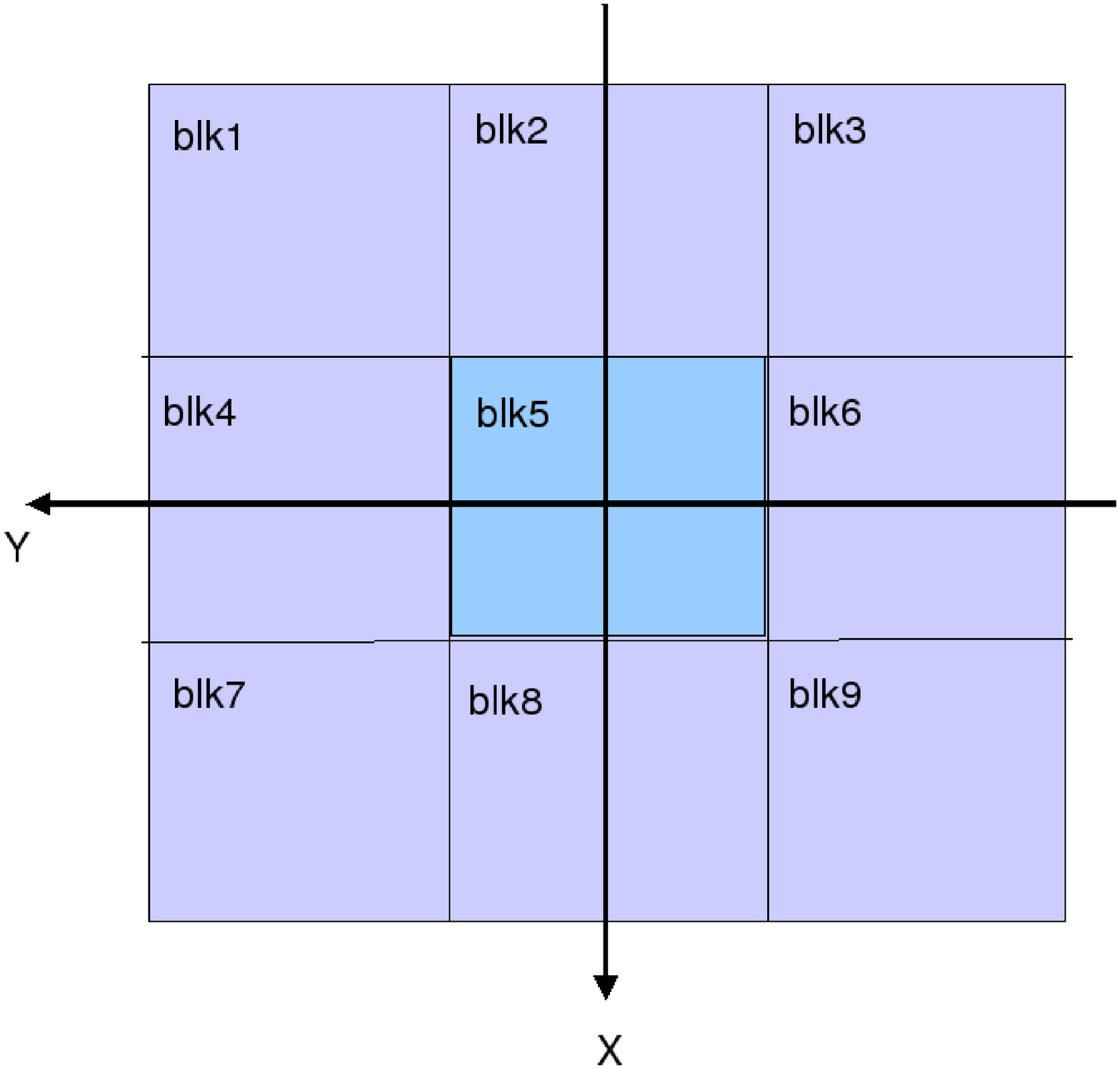}\\
      Fig.~1 \quad 9 block scheme for NaI (Tl) calibration
    \end{center}
   The amplitude ratios of adjacent blocks to the central one ($|x|<0.03m, |y|<0.03m$) were fitted to 2nd-order polynomials in two dimensions.
   The following relations were obtained:
    \begin{equation}
R1=0.01473-0.09517x+7.359x^2+0.1848y+7.51y^2\,, 
    \end{equation}
   \begin{equation}
R2=0.01402-0.3942x+8.537x^2-0.001413y-2.994y^2 \,,
    \end{equation} 
     \begin{equation}
R3=0.01399-0.1175x+4.114x^2-0.1725y+4.55y^2 \,,
    \end{equation} 
 \begin{equation}
R4=0.02414-0.08904x+21.29x^2+0.8816y+23.54y^2\,, 
    \end{equation} 
    \begin{equation}
    R5=1 \,,
    \end{equation}  
    \begin{equation}
R6=0.02423-0.001044x+19.357x^2-0.5776y+9.302y^2 \,,
    \end{equation}  
 \begin{equation}
R7=0.008895-0.09234x+6.1917x^2+0.09544y+4.266y^2 \,,
    \end{equation} 
   \begin{equation}
R8=0.01739+0.7615x+22.78x^2+0.01578y+5.965y^2 \,,
    \end{equation} 
     \begin{equation}
R9=0.01359+0.1170x+7.258x^2-0.2098y+7.97y^2 \,,
    \end{equation} 
 where R1, R2, R3, R4, R5, R6, R7, R8 and R9 are the amplitude ratios of each adjacent block to the central block, respectively. The $x$ and $y$ represent the vertical and horizontal directions shown in Fig. 1  Since the amplitude ratios of adjacent blocks has a momentum dependence, different math forms were taken for different momentum settings for which calibrations were performed.
    \section{Checking the calibration results }
    The energy deposition $E$ of an incident electron in the calorimeter detector could be calculated with the calibration constants $C$ by the following formula:
  \begin{center}
      \begin{equation}
        E=\sum_{i=0}^{90}C_{i}\cdot A_{i}\,,
      \end{equation}
    \end{center}
 \vspace{1mm}
where $i$ was the number of NaI (Tl) detector block. The $E/p$ of electrons should be around 1 after the calibration correction where $E$ was the total deposit energy calculated from Eq. (19) and $p$ was the momentum. The $E/p$ plot before calibration is shown in Fig. 2 at 120 MeV of scattering electron. The plot after calibration correction is shown in Fig. 3.
    \begin{center}
      \includegraphics[width=8cm,height=5cm]{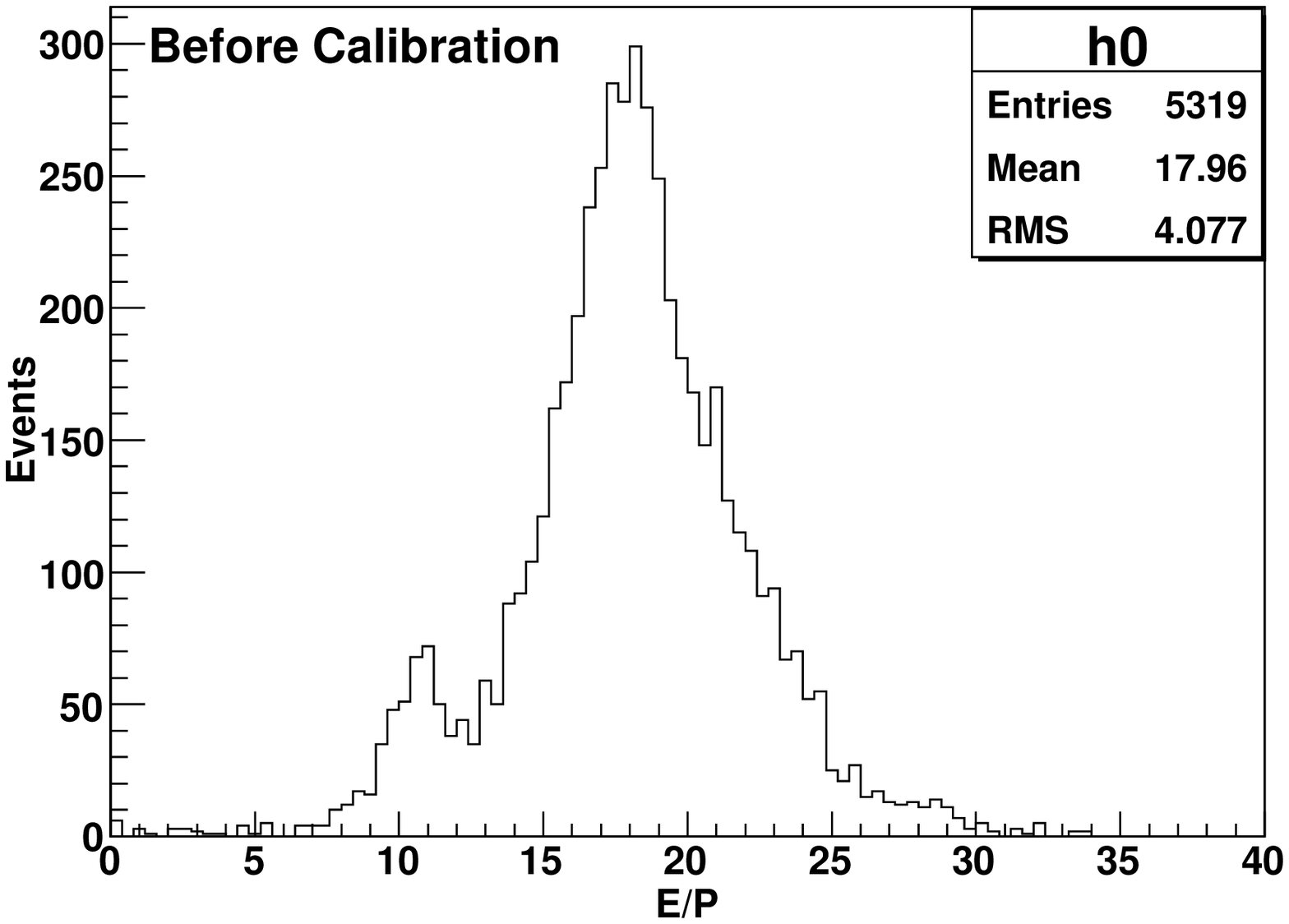}\\
      Fig.~2 \quad  The E/p plot before calibration for a non-calibration run
    \end{center}
    The E/p plot for electrons should be a Gaussian distribution plus a tail. The width of the Gaussian distribution represents the detector's resolution. Fig. 3 shows a reasonable spectrum after calibration correction.
  \begin{center}
      \includegraphics[width=8cm,height=5cm]{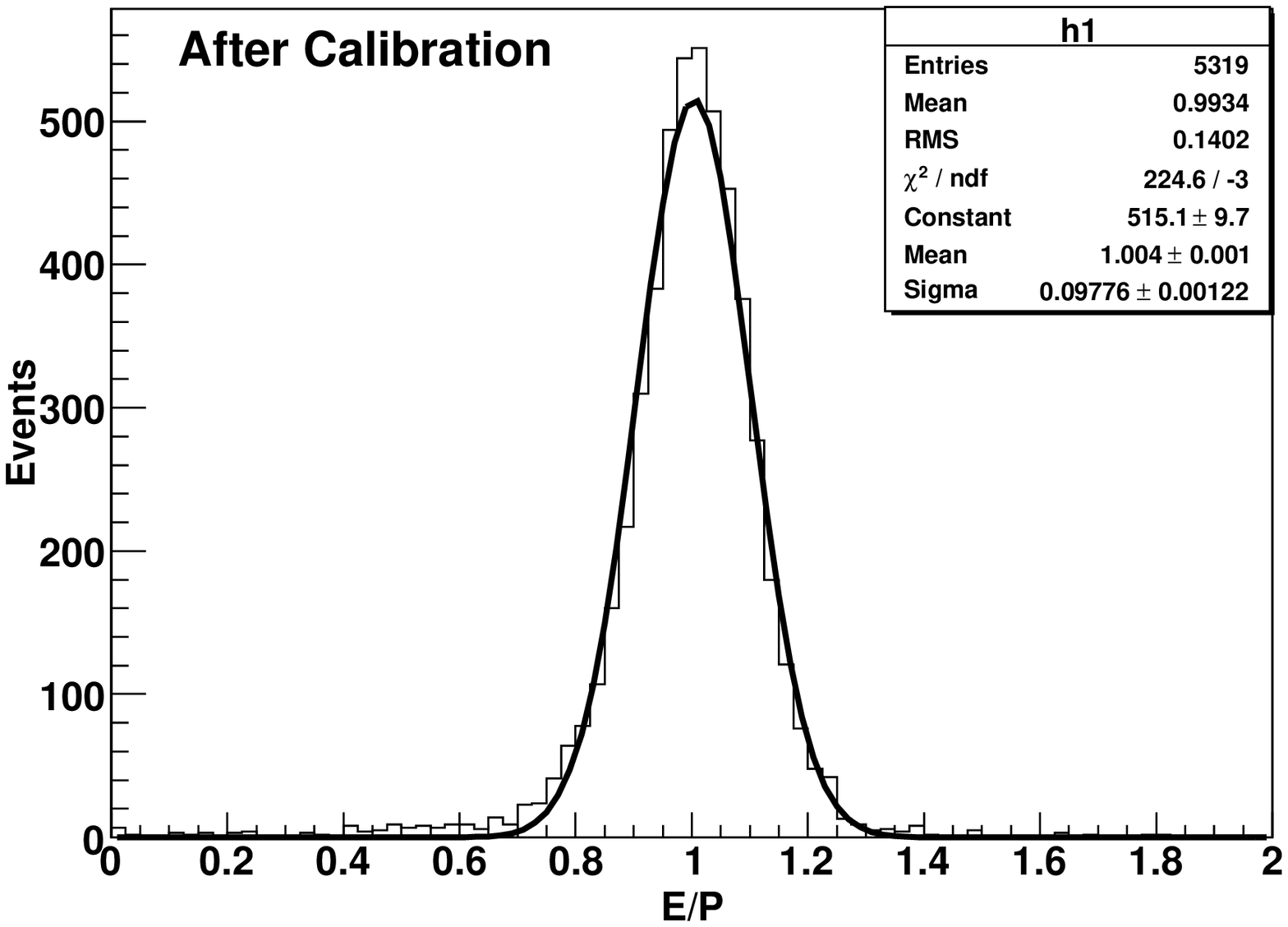}\\
      Fig.~3 \quad  E/p plot after calibration for the electron peak of a non-calibration run   
    \end{center}
The events at the low momentum tail were electrons with energy leakage, electrons that scattered  the walls of the spectrometer and secondary productions, including electrons and hadrons.
   The energy resolution of E/P in Fig. 3 is about $9.7\%$ for 120 MeV at this HV setting and the best resolution we can reach is $3\%$ for 1 GeV. Due to a large set of kinematics and the changes of the HV for NaI (Tl) detector during the data-taking period, different sets of calibration constants were needed for this experiment. A total of 40 sets of constants for production runs were obtained for this experiment.
    
\section{NaI (Tl) simulation for background analysis}
    A simulation using SNAKE\cite{snake} and GEANT3\cite{GEANT3} was performed when the experiment was proposed\cite{pro}. We used that to generate an input electron sample for the NaI (Tl) detector. GEANT4 was used for NaI (Tl) detector simulation.  
    \subsection{SNAKE and GEANT3 simulation}
  The background generated by the interaction of electrons with the inner walls of the spectrometer magnets was studied with a Monte-Carlo simulation. The simulation is based on a ray-tracing program, SNAKE. In the original version of SNAKE, an electron hitting the internal boundaries of the spectrometer was considered lost. In the modified version of the simulation program, the electron was studied further with a GEANT3 simulation for one of two possibilities: (a) scattering off the wall; (b) generation of secondary particles from an interaction with the wall material. 
\begin{center}
\vspace{1mm} 
\includegraphics[width=8cm,height=5cm]{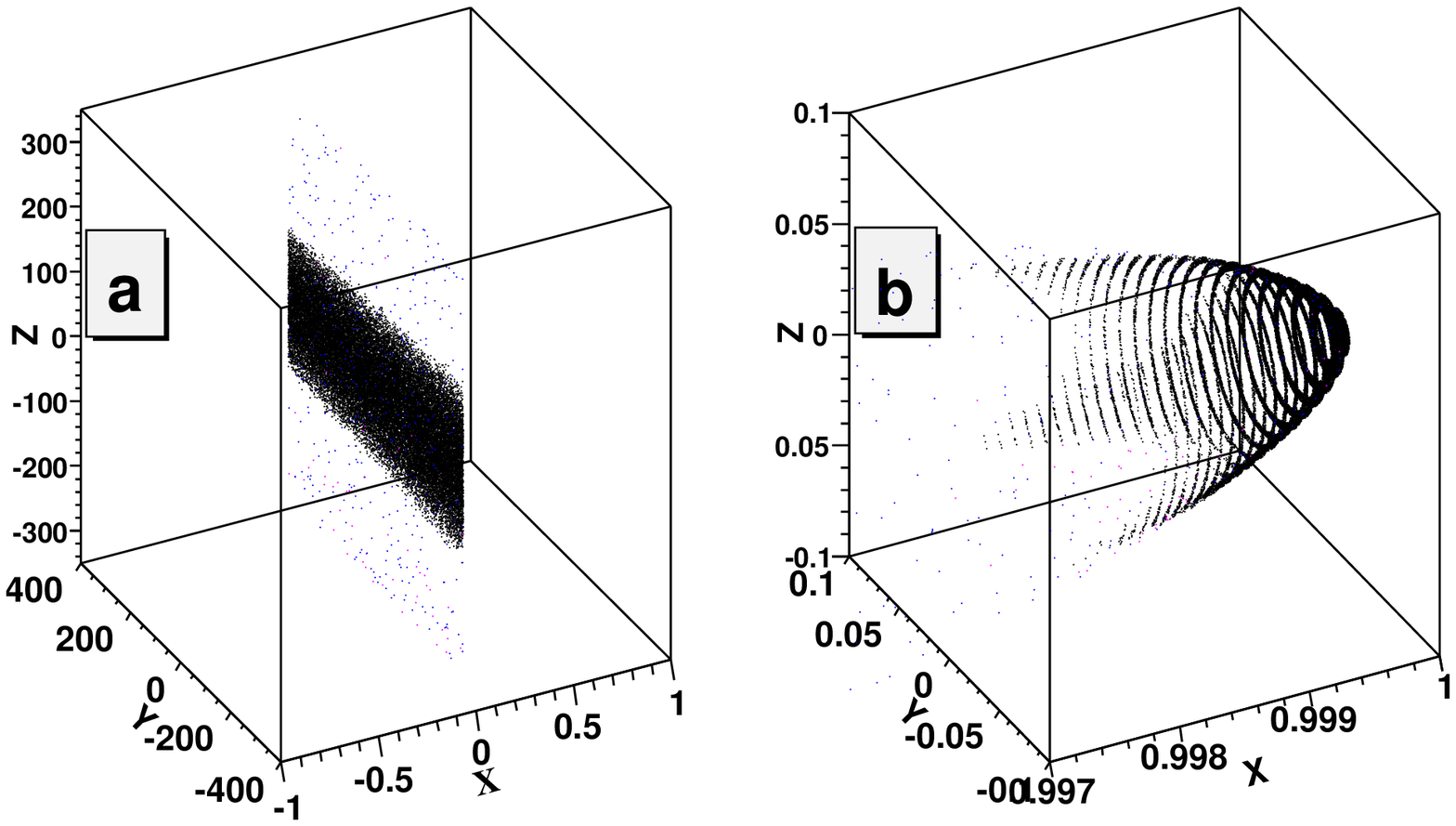}\\
 Fig.~4 \quad The position and direction of  electron sample generated by SNAKE and GEANT3 at a momentum setting of 120 MeV\\
\end{center}
Then the rescattered electron or the secondary particles were re-inserted into the SNAKE simulation and were traced to the focal plane. Since particles hitting the walls of Q1, Q2 could not make to the focal plane and the ones hitting the dipole have small probabilities to reach the focal plane in the SNAKE simulations. So the simulation was focused on the  interaction of electrons with the walls of the Q3 magnet. Due to the proximity of the Q3 magnet to the focal plane, electrons bouncing off the surface of the Q3 magnet would have a higher probability of survival than those bounced off the other magnets. The result of simulation shows that the background generated in this process is about 1.2$\%$ of the clean events at a spectrometer momentum setting of 120 MeV/c, the lowest momentum setting among our kinematics. A few background events with energy comparable to clean events came from a single, large angle scattering on the surface of the Q3 magnet. With a tight cut on the position on the focal plane, about 80$\%$ of the background events were eliminated. This is in agreement with results from an independent analysis\cite{Ja}. The remaining background events after the focal plane position cut can be eliminated by an independent energy measurement such as NaI (Tl) calorimeter. The position and momentum direction of electron sample before entering NaI (Tl) detector which were generated from the SNAKE and GEANT3 simulation were shown in the Fig. 4 (a) and (b), respectively. The electrons reflected by the Dipole and Q3 contributed about 0.24$\%$, 1.2$\%$ background, respectively. 
\begin{center}
\vspace{1mm}
\includegraphics[width=8cm,height=5cm]{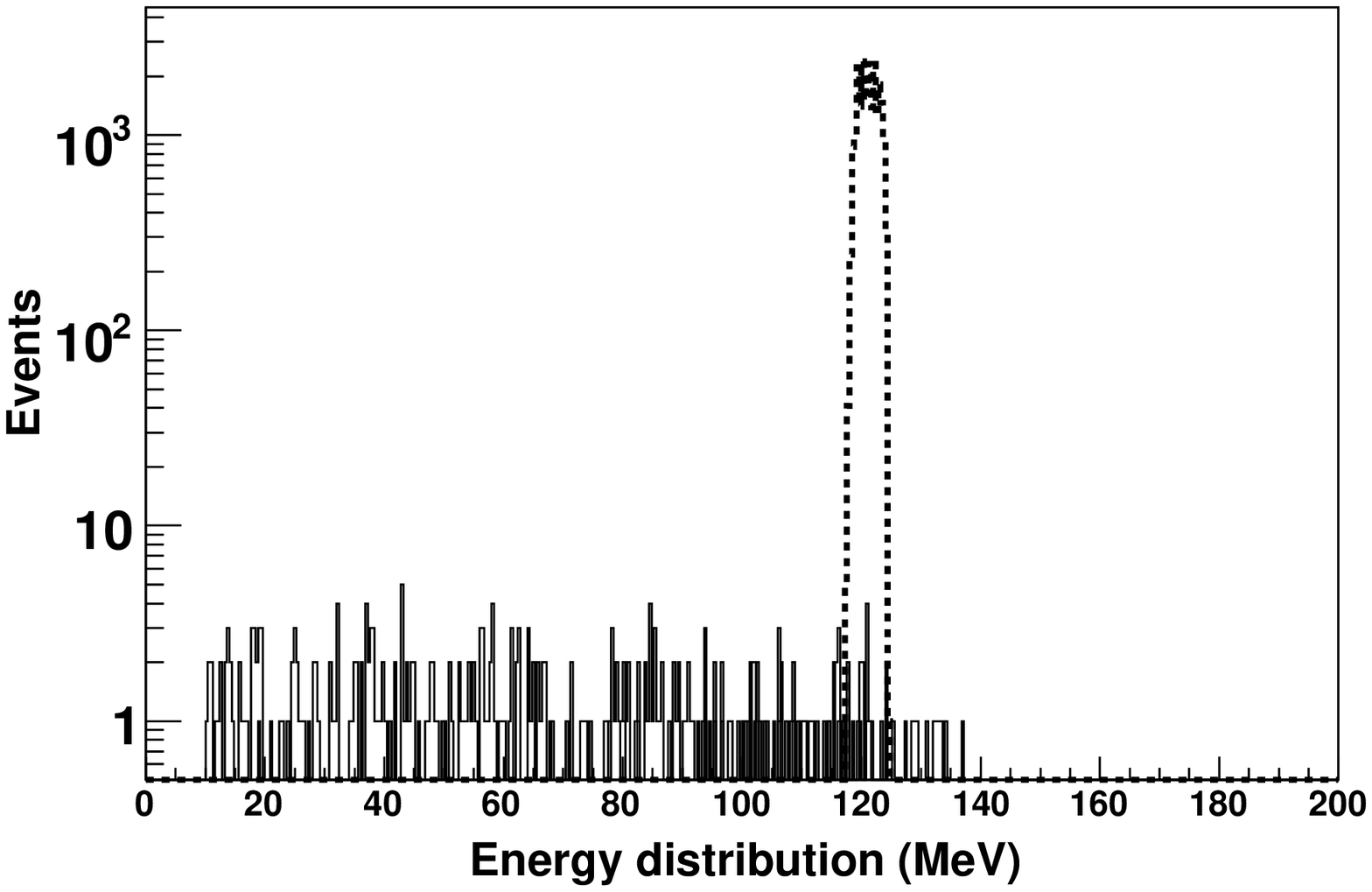}\\
 Fig.~5 \quad The energy distribution of  electron sample generated by SNAKE and GEANT3 at a momentum setting of 120 MeV
\end{center}
The fig. 5 shows the energy distribution for the electron sample. The good electron peak (dash line) is at 120 MeV and the background (solid line) spreads out, which are electrons scattered by Dipole and Q3 magnets. The electrons were re-inserted into GEANT4 to simulate their behavior in the NaI (Tl) detector.
 
\subsection{NaI (Tl) GEANT4 simulation}
The geometry of the NaI (Tl) detector is shown in Fig. 6. The parameters used in the GEANT4 simulation for NaI (Tl) properties were obtained from the
manufacture\cite{saint} . The trajectories of electrons and photons are shown in Fig. 6 . Since the input electron energy was 120 MeV, the total energy of the electron was absorbed by NaI (Tl) calorimeter.
\begin{center}
\vspace{1mm}
\includegraphics[width=8cm,height=8cm]{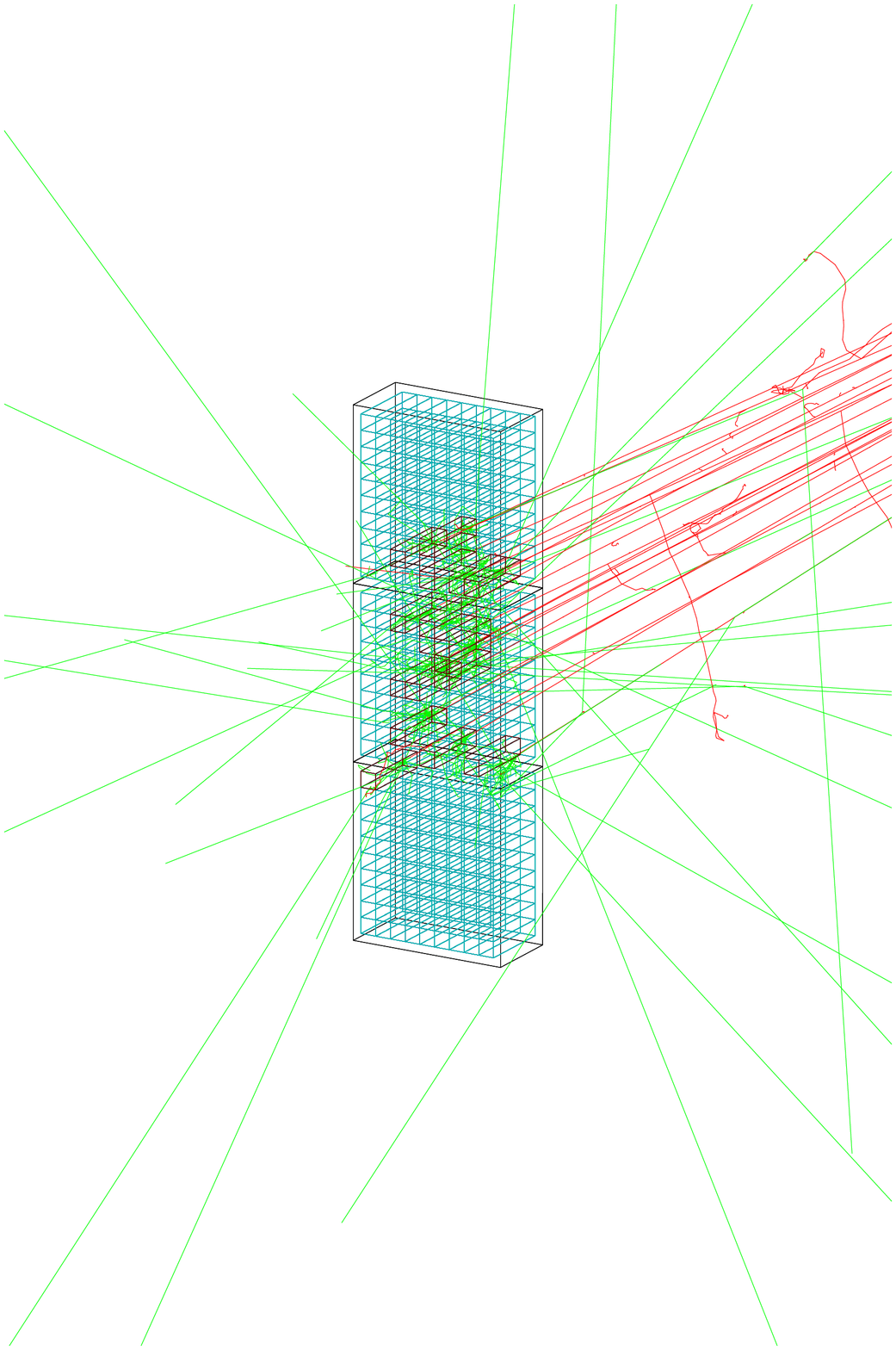}\\
 Fig.~6 \quad Schematic of GEANT4 simulation for NaI (Tl) detector at a momentum setting of 120 MeV of scattering electron; Red line (electron), Green line (photon).
\end{center}

\subsection{A comparison between data and simulation for 120 MeV and 539 MeV}
Fig. 7 shows, in log scale, the distribution of energy deposition by electrons in the NaI (Tl) detector for simulation result(dash line) and data(solid line), respectively. Since the simulation did not contain spectrometer intrinsic resolution, hence the simulation only reproduces the low energy part of Gaussian distribution after electron deposited energy in NaI (Tl) middle box. Becasuse the low energy background part was the main issue, in our case the asymmetry of simulation result has no effect on the background analysis.

\begin{center}
\vspace{1mm}
\includegraphics[width=8cm,height=5cm]{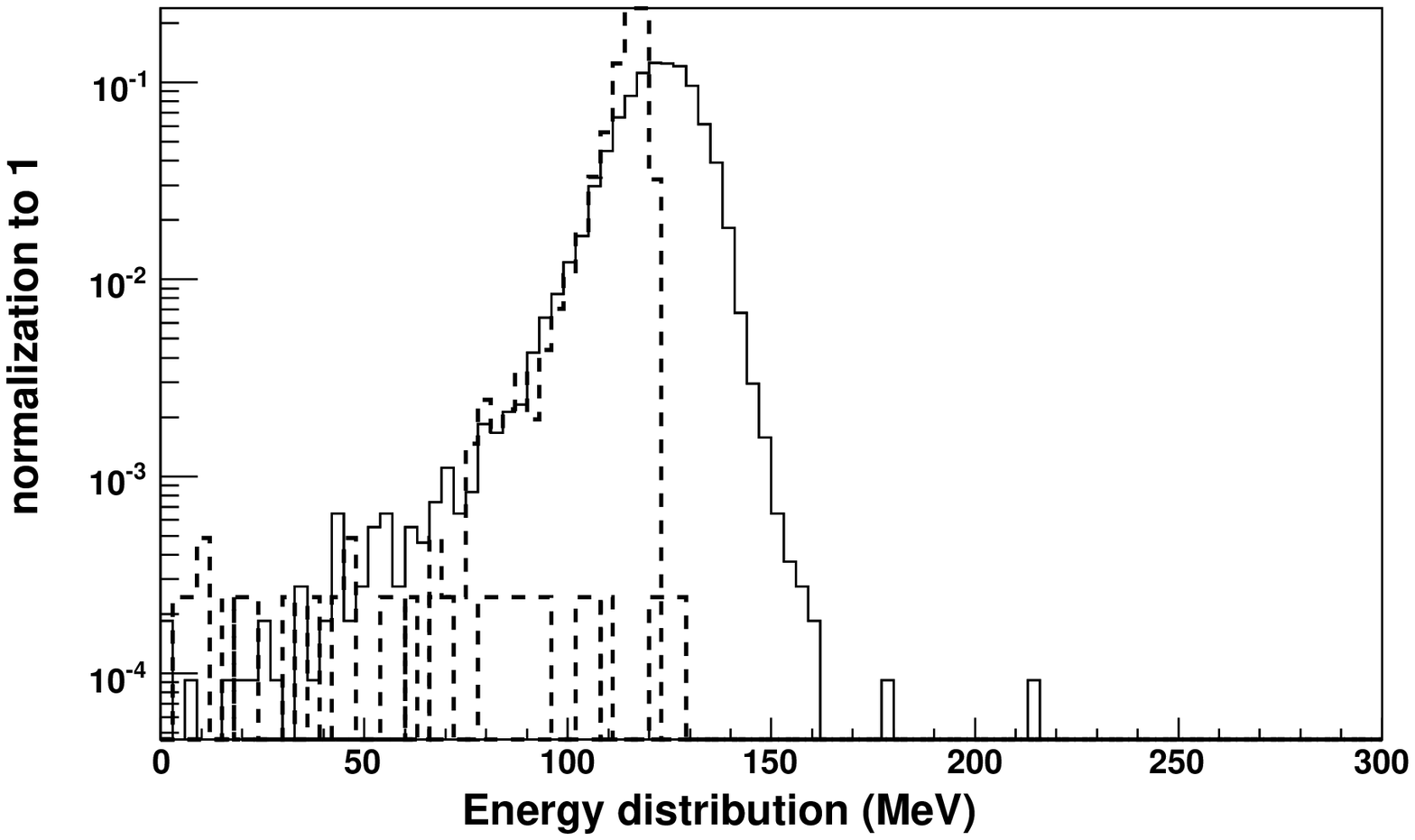}\\
 Fig.~7 \quad Comparison between data (solid line) and simulation (dash line) at a momentum setting of 120 MeV
\end{center}
 The good match between simulation and real data in low energy part of distribution was shown in Fig. 7 for 120 MeV of scattering electrons. From simulation result, the contamination from scattering off walls of the Dipole and Q3 was about 0.31$\%$ and good electron cut efficiency was 99.9$\%$ after adding a cut on the energy distribution at 50 MeV for this kinematic setting.
\begin{center}
\vspace{1mm}
\includegraphics[width=8cm,height=5cm]{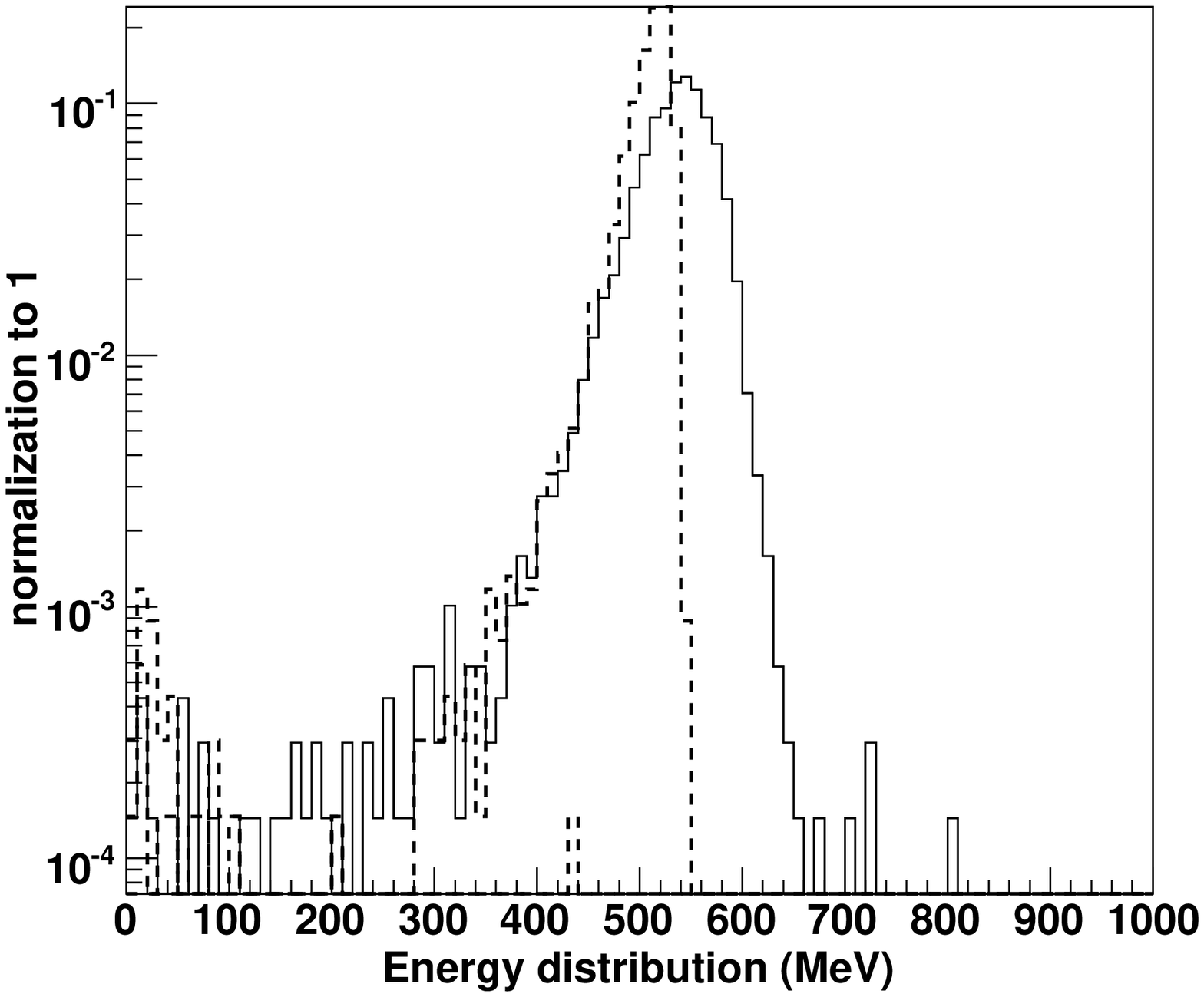}\\
 Fig.~8 \quad Comparison between data (solid line) and simulation (dash line) for 539 MeV
\end{center}
  The fig. 8 shows a comparison between the simulation and data for scattering electrons with momenta of 539 MeV. With a cut applied at 150 MeV, the residual contamination from surface scattering from the Dipole and Q3 was about 0.029$\%$ and good electron cut efficiency was 99.9$\%$. 

\section{Conclusion}
  The NaI (Tl) detector has been calibrated for the Coulomb Sum Rule experiment. Corrections were applied to the missing energy due to a few inefficient detector blocks.  Because of the large set of kinematics and the changes in HVs, 40 sets of calibration constants were obtained. The energy resolution of the NaI (Tl) detector reached $\frac{\delta E}{\sqrt{E}}\approx 3\%$ for 1 GeV electrons. We also did simulation to study the background due to re-scattering from the inner walls of the Dipole and Q3 for the spectrometer momentum settings of 120 MeV and  539 MeV. The contamination was about 0.3$\%$ and 0.03$\%$ when cuts at 50 MeV and 150 MeV were applied for the momentum settings of 120 MeV and 539 MeV, respectively (see Fig. 7 and Fig. 8). With the same cuts, good electron cut efficiency was 99.9$\%$ for both settings.\\

\end{multicols}

\vspace{10mm}
\vspace{-1mm}
\centerline{\rule{80mm}{0.1pt}}
\vspace{2mm}

\begin{multicols}{2}

\end{multicols}

\clearpage


\begin{thebibliography}{90}

\vspace{3mm}

\bibitem{morgenstern}  J.Morgenstern, Z.E. Meziani. Phys. Lett. B, 2001 515:269-275

\bibitem{pro}  J.P. Chen, S. Choi, Z.E. Meziani (Spokespersons). Jefferson Lab E05-110

\bibitem{Al}   J.Alcorn et al. Nuclear Instruments and Methods in Physics Research A, 2004, 522:294-346

\bibitem{NaINIM} S.L. WILSON et al. Nuclear Instruments and Methods in Physics Research, 1988, A264

\bibitem{zhu} Zhu R Y. Nuclear Instruments and Methods in Physics Research A:Accelerators, Spectrometers, Detectors and Associated Equipment, 1998, 413:297-311

\bibitem{snake} http://hallaweb.jlab.org/news/minutes/tranferfuncs.html

\bibitem{GEANT3}  http://www.pv.infn.it/sc/cern/geant.pdf

\bibitem{Ja} J.Arrington.  Presentation given at JLab E01-001 collaboration meeting, 2003

\bibitem{saint}  http://www.detectors.saint-gobain.comi/NaI (Tl).aspx




\end{thebibliography}
\end{document}